\documentclass[12pt]{article}
\usepackage{graphicx}
\catcode`\@=11
\def\numberbysection{\@addtoreset{equation}{section}
\def\theequation{\thesection.\arabic{equation}}}

\numberbysection

\newcommand{\beq}{\begin{equation}}
\newcommand{\beqa}{\begin{eqnarray}}
\newcommand{\eeq}{\end{equation}}
\newcommand{\eeqa}{\end{eqnarray}}
\newcommand{\abs}[1]{\vert#1\vert}
\newcommand{\bigmean}[1]{\left\langle#1\right\rangle}
\renewcommand{\c}{{\rm c}}
\newcommand{\cum}[1]{\langle\!\langle#1\rangle\!\rangle}
\renewcommand{\d}{{\rm d}}
\newcommand{\demi}{{\textstyle\frac12}}
\newcommand{\e}{{\rm e}}
\newcommand{\eps}{\varepsilon}
\newcommand{\frad}[2]{\displaystyle{\displaystyle#1\over\displaystyle#2}}
\newcommand{\g}{\gamma}
\renewcommand{\i}{{\rm i}}
\newcommand{\mean}[1]{\langle#1\rangle}
\newcommand{\quart}{{\textstyle\frac14}}
\newcommand{\rms}{{\rm rms}}
\newcommand{\tquart}{{\textstyle\frac34}}
\newcommand{\x}{{\rm cr}}
\newcommand{\var}{\mathop{{\rm var}}}
\newcommand{\C}{{\bf C}}
\newcommand{\D}{\Delta}
\newcommand{\E}{{\bf E}}
\newcommand{\K}{{\bf K}}
\renewcommand{\Re}{\mathop{{\rm Re}}}

\begin{document}
\centerline{\Large\bf On the statistics of superlocalized states}
\vspace{.3cm}
\centerline{\Large\bf in self-affine disordered potentials}
\vspace{1.6cm}
\centerline{\large J.M.~Luck\footnote{luck@spht.saclay.cea.fr}}
\vspace{.4cm}
\centerline{Service de Physique Th\'eorique\footnote{URA 2306 of CNRS},
CEA Saclay, 91191 Gif-sur-Yvette cedex, France}
\vspace{1cm}
\begin{abstract}
We investigate the statistics of eigenstates
in a weak self-affine disordered potential in one dimension,
whose Gaussian fluctuations grow with distance
with a positive Hurst exponent $H$.
Typical eigenstates are superlocalized on samples much larger than
a well-defined crossover length, which diverges in the weak-disorder regime.
We present a parallel analytical investigation
of the statistics of these superlocalized states in
the discrete and the continuum formalisms.
For the discrete tight-binding model, the effective localization length
decays logarithmically with the sample size,
and the logarithm of the transmission is marginally self-averaging.
For the continuum Schr\"odinger equation,
the superlocalization phenomenon has more drastic effects.
The effective localization length decays as a power of the sample length,
and the logarithm of the transmission is fully non-self-averaging.
\end{abstract}
\vfill
\noindent To be submitted for publication to the Journal of Physics A

\noindent P.A.C.S.: 71.23.An, 73.20.Fz, 72.15.Rn, 05.40.-a.

\newpage
\section{Introduction}

The Anderson localization of a quantum-mechanical particle
by a random potential is by now an old and well-understood problem~\cite{loc}.
This is especially so in one dimension.
Consider for definiteness the tight-binding equation:
\beq
\psi_{n+1}+\psi_{n-1}+V_n\psi_n=E\psi_n.
\label{tb}
\eeq
In the usual situation where the site potentials $V_n$ are uncorrelated,
all the eigenstates are known to be exponentially localized
(see e.g.~\cite{thou,pen}).

The peculiar features of the localization problem
for various kinds of one-dimensional disordered potentials
with non-trivial correlations have also been investigated,
including dimer models~\cite{dim},
potentials with power-law correlations~\cite{pow},
potentials generated by chaotic dynamical systems~\cite{cha},
and potentials whose correlations are designed on purpose~\cite{adh}.
These examples share the common feature of {\it spatial stationarity}:
the statistics of the potential is invariant under translation, so that
its two-point correlation has a well-defined thermodynamical limit
$\mean{V_mV_n}=C_{m-n}$, which only depends on the distance $\abs{m-n}$.

The more exotic situation of {\it non-stationary} random potentials,
whose fluctuations grow with distance,
has only been considered more recently~\cite{m1,m2,b1,m3,b2,b3}.
An example of most physical interest is that of a
{\it self-affine} Gaussian potential with Hurst exponent $0<H<1$, such that
\beq
\mean{(V_m-V_n)^2}=\D^2\abs{m-n}^{2H}.
\label{hur}
\eeq
Such a sequence of potentials can be generated
by fractional Brownian motion,
with a proper choice of stationary but correlated increments
\beq
\eps_n=V_n-V_{n-1}.
\label{inc}
\eeq
The particular case of the usual Brownian motion,
corresponding to stationary and independent increments
($\mean{\eps_m\eps_n}=\D^2\delta_{mn}$), has a Hurst exponent $H=1/2$.

An alternative way of characterizing non-stationary potentials
consists in considering their structure factor
$S(q)=\mean{\hat V(q)\hat V(-q)}$,
assuming a power-law divergence of the form
\beq
S(q)\sim\abs{q}^{-\alpha}
\eeq
in the long-wavelength limit ($q\to0$).
Only the smaller values of the scaling exponent ($\alpha<1$)
correspond to stationary potentials,
with long-range correlations falling off as $C_n\sim\abs{n}^{-(1-\alpha)}$.
Larger values of $\alpha$ necessarily correspond to non-stationary potentials.
The above self-affine potentials with stationary increments
are obtained in the range $1<\alpha<3$, with the correspondence $\alpha=2H+1$.

The main novel feature induced by the non-stationarity of the potential
is that the effective disorder strength depends on the spatial scale.
The typical potential fluctuation over a distance $N$
indeed grows as $V_\rms(N)=\D N^H$.
It becomes therefore comparable with the bandwidth when the distance
reaches the crossover length
\beq
N_\x=\D^{-1/H}.
\label{nx}
\eeq
The situation of most interest is that of a weak disorder ($\D\ll1$),
so that $N_\x\gg1$.
The eigenstates are conventionally localized on scales $N\ll N_\x$,
with a localization length scaling as $\xi\sim1/\D^2$.
On larger scales ($N\gg N_\x$), however,
eigenstates become strongly localized or {\it superlocalized}~\cite{b2,b3},
because wavefunctions fall off very fast
in the classically forbidden zones of the potential ($\abs{E-V_n}>2$).
Let us mention that an alternative viewpoint~\cite{m1,m2}
consists in keeping fixed the effective potential width
$V_\rms(N)=\D N^H=\Sigma$ at the scale of the system size~$N$.
The microscopic disorder strength $\D=\Sigma/N^H$
therefore gets rescaled to smaller and smaller values.
The crossover length $N_\x=N/\Sigma^{1/H}$
grows proportionally to the sample size,
whereas the localization length scales as $\xi\sim N^{2H}$.
For a large enough non-stationarity ($H>1/2$),
such that $\xi$ may become much larger than the sample size $N$,
a crossover line in the $(\Sigma,E)$ plane
between extended and localized states
has been observed~\cite{m1,m2}, and theoretically explained~\cite{b2,b3}.

The goal of the present work is to provide the first quantitative analysis
of the statistics of superlocalized eigenstates,
especially in the regime $N\gg N_\x$,
where the superlocalization phenomenon is fully developed.
We shall successively deal with the discrete tight-binding model (Section~2)
and the continuum Schr\"odinger equation (Section~3).

The key quantity considered throughout this work
is the effective Lyapunov exponent
\beq
\g_N=\frac{1}{N}\ln\abs{\psi_N},
\label{ly}
\eeq
where $\psi_n$ is the generic (growing) solution to~(\ref{tb}).
The effective Lyapunov exponent provides an estimate of the global growth rate
of this solution over $N$ lattice sites,
and therefore of the effective decay rate of eigenstates over the same range.
In other words, the effective localization length at scale $N$ is
\beq
\xi_N=\frac{1}{\g_N}.
\eeq
The effective Lyapunov exponent is also a central quantity in
the theory of coherent quantum transport.
Indeed the two-probe Landauer formula~\cite{lan}
expresses the zero-temperature conductance $g_N$
of a sample made of $N$ lattice sites,
in terms of the intensity transmission $T_N$ across that sample, as
\beq
g_N=\frac{2e^2}{h}\,T_N.
\label{lanfor}
\eeq
Furthermore, in the insulating regime where the transmission is small,
it is known to scale as $T_N\sim 1/\abs{\psi_N}^2$, hence
\beq
\ln T_N\approx-2N\g_N.
\label{lnt}
\eeq

The effective Lyapunov exponent,
and related physical quantities such as the effective localization length
and the logarithm of the transmission (conductance),
exhibit different kinds of scaling behavior
in the two situations to be investigated successively,
the discrete tight-binding model (Section~2)
and the continuum Schr\"odinger equation (Section~3).
This basic difference is further commented on in the Discussion (Section~4).

\section{The discrete tight-binding model}

\subsection{Reminder}

We find it useful to first give a brief reminder on
the conventional situation of a stationary random potential.
The statistics of the size-dependent effective Lyapunov exponent $\g_N$
is then very similar to that of the free-energy density
of a disordered statistical-mechanical system~\cite{pen,cpv,alea}.
The effective Lyapunov exponent $\g_N$ has a well-defined self-averaging limit
$\g$ in the $N\to\infty$ limit, simply referred to as the Lyapunov exponent.
Its reciprocal $\xi=1/\gamma$ is interpreted as the localization length
of the problem.

The fluctuations of $\g_N$ around $\g$ are Gaussian and scale as
$\var{\g_N}=\mean{(\g_N-\g)^2}\approx\sigma^2/N$
for a finite but large enough sample ($N\g\gg1$).
More generally, the product $N\g_N$ is extensive,
in the strong sense that all its cumulants
\beq
\cum{(N\g_N)^k}\approx c_k N
\eeq
grow linearly with $N$~\cite{pen}, with amplitudes $c_1=\g$, $c_2=\sigma^2$,
and so on.
It is worth noticing that $N\g_N$, which is analogous to the total free energy,
is also a physical observable in the present context,
as it is nothing but the logarithm of the transmission (conductance),
up to a constant factor [see~(\ref{lnt})].

Furthermore, in the usual situation of independent site potentials
with $\mean{V_n}=0$ and $\mean{V_mV_n}=W^2\delta_{mn}$,
the Lyapunov exponent behaves as follows in the weak-disorder limit ($W^2\ll1$):

\begin{itemize}

\item Inside the band, i.e., for $\abs{E}<2$, setting $E=2\cos q$,
the celebrated result~\cite{thou}
\beq
\g\approx\frad{W^2}{8\sin^2q}=\frad{W^2}{2(4-E^2)}
\label{in}
\eeq
shows that the localization length diverges as $1/W^2$.

\item Outside the band, i.e., for $\abs{E}>2$, setting $\abs{E}=2\cosh t$,
i.e.,
\beq
t=\ln\frad{\abs{E}+\sqrt{E^2-4}}{2}>0,
\eeq
the generic solution to~(\ref{tb}) grows exponentially as $\psi_n\sim\e^{nt}$
in the absence of disordered potential, hence
\beq
\g\to t\qquad(W^2\to0).
\label{out}
\eeq

\item
Near band edges, i.e., for $E\to\pm2$ and $W^2\to0$ simultaneously,
the Lyapunov exponent obeys a scaling law of the form
\beq
\g\approx W^{2/3}\;F\!\left(\frad{\abs{E}-2}{W^{4/3}}\right),
\eeq
where the scaling function $F$ is explicitly known
in terms of Airy functions~\cite{dg}.

\end{itemize}

\subsection{General results}

We now turn to the current problem,
namely the tight-binding equation~(\ref{tb}) with a weak self-affine potential
whose fluctuations grow with distance according to~(\ref{hur}),
with $\D^2\ll1$.
We choose the zero of energies as the site potential at the origin ($V_0=0$),
and assume that the origin sits in an allowed zone, i.e., $\abs{E}<2$.

A long enough sample ($N\gg N_\x$) generically
consists of an alternation of allowed zones
($\abs{E-V_n}<2$) and of forbidden zones ($\abs{E-V_n}>2$).
It can be argued from the above results
that the growth rate of a generic wavefunction
is proportional to $\D^2$ in allowed zones,
i.e., inside the `local band' around site $n$
(where energies are shifted by $V_n$)~[see~(\ref{in})],
whereas it is of order unity in forbidden zones,
i.e., outside the `local band'~[see~(\ref{out})].

This picture is fully corroborated by the plots shown in
References~\cite{b2,b3},
where eigenstates are indeed seen to be essentially constant
in allowed zones, and to drop very suddenly in forbidden zones.
Our main goal is to turn the above picture
into a quantitative description of the statistics
of the effective Lyapunov exponent~(\ref{ly}).

Our starting point is the approximate formula
\beq
\psi_N\approx\exp\left(\sum_{n=0}^{N-1} t_n\right),
\label{wd}
\eeq
with the definition
\beq
E-V_n=2\cosh t_n,
\eeq
hence
\beq
\g_N=\frac{1}{N}\sum_{n=0}^{N-1}\Re t_n,
\label{lyd}
\eeq
with
\beq
\Re t_n=\left\{\matrix{
\ln\frad{\abs{E-V_n}+\sqrt{(E-V_n)^2-4}}{2}>0\hfill
&\hbox{in forbidden zones}&(\abs{E-V_n}>2),\hfill\cr\cr
0\hfill&\hbox{in allowed zones}\hfill&(\abs{E-V_n}<2).\hfill
}\right.
\label{dichod}
\eeq

Let us first comment on~(\ref{wd}) for a while.
This expression is a full discrete analogue of the W.K.B.
integral~(\ref{psiwkb}), to be used in Section~3.
It provides a quantitative description of the growing solution to~(\ref{tb})
in the forbidden zones of the potential,
under the sole hypothesis that the sequence
of site potentials has small increments:
\beq
\eps_n=V_n-V_{n-1}\ll1.
\eeq
Expression~(\ref{wd}) can be easily derived by introducing
the Riccati variables~\cite{alea,tn,prb}
\beq
R_n=\frac{\psi_{n+1}}{\psi_n}.
\label{ric}
\eeq
Assuming $\psi_0=1$, we have $\psi_n=R_0\cdots R_{n-1}$, and
\beq
\g_N=\frac{1}{N}\sum_{n=0}^{N-1}\ln\abs{R_n}.
\label{garic}
\eeq
The variables $R_n$ obey the recursion
\beq
R_n=E-V_n-\frac{1}{R_{n-1}}.
\label{ricrec}
\eeq
If we now set
\beq
x_n=\frad{\e^{t_n}-R_n}{1-\e^{t_n}R_n},\qquad
R_n=\frac{\e^{t_n}-x_n}{1-\e^{t_n}x_n},
\eeq
the variables $x_n$ obey
\beq
x_n=\e^{-2t_n}\frac{x_{n-1}+\delta_n}{1+\delta_nx_{n-1}},
\label{xrec}
\eeq
with
\beq
\delta_n=\frac{\e^{t_{n-1}}-\e^{t_n}}{\e^{t_n+t_{n-1}}-1}
\approx\frac{\eps_n}{4\sinh^2 t_n},
\eeq
to leading order as $\eps_n\ll1$.
To the same order, the recursion~(\ref{xrec}) can be linearized to
$x_n\approx\e^{-2t_n}(x_{n-1}+\delta_n)$,
implying that the $x_n$'s are proportional to the $\eps_n$'s.
One has therefore $R_n\approx\exp(t_n)$,
up to terms proportional to the $\eps_n$'s.
The expressions~(\ref{wd}) and~(\ref{lyd}) follow at once.
Keeping track of higher powers of the $\eps_n$'s in the above equations
would be an efficient way of performing systematic
weak-disorder expansions~\cite{alea,prb}.

Our analysis of the statistics of the effective Lyapunov exponent $\g_N$
is based on the estimate~(\ref{lyd}).
It turns out to be advantageous to introduce the Laplace representation
\beq
\Re t_n=\int\!\frac{\d s}{\i\pi s}\,\cosh(sE)\,K_0(2s)\,\exp(sV_n),
\eeq
where $K_0$ is the modified Bessel function,
and the integration contour is a vertical line with $\Re s>0$.
Using the Gaussian statistics of the potentials $V_n$,
with $V_0=0$ and~(\ref{hur}), we have
\beq
\mean{\exp(sV_n)}=\exp\left(\demi\D^2s^2n^{2H}\right),
\eeq
and similar expressions for higher-order characteristic functions
such as $\mean{\exp(s_1V_m+s_2V_n)}$.
We can therefore express the correlation functions of the variables $\Re t_n$
as multiple contour integrals.
By means of~(\ref{lyd}), the moments of the effective Lyapunov exponent $\g_N$
can, in turn, be expressed as multiple integrals.
The rescaling $x=m/N_\x$, $y=n/N_\x$ implies that, as could be expected,
the final results only depend on the system size $N$
through the dimensionless ratio
\beq
X=\frac{N}{N_\x}=\D^{1/H}N,
\eeq
where the crossover scale $N_\x$ has been introduced in~(\ref{nx}).
We thus obtain
\beqa
&&\mean{\g_N}=\frac{1}{X}\int_0^X\d x
\int\!\frac{\d s}{\i\pi s}\,\cosh(sE)\,K_0(2s)\,
\exp\left(\demi s^2x^{2H}\right),\label{conave}\\
&&\mean{\g_N^2}=\frac{1}{X^2}\int_0^X\d x\int_0^X\d y
\int\!\frac{\d s_1}{\i\pi s_1}\,\cosh(s_1E)\,K_0(2s_1)\,
\int\!\frac{\d s_2}{\i\pi s_2}\,\cosh(s_2E)\,K_0(2s_2)\,\nonumber\\
&&{\hskip 20truemm}\times
\exp\left(\demi\!\left(s_1^2x^{2H}+s_1s_2(x^{2H}+y^{2H}-\abs{x-y}^{2H})
+s_2^2y^{2H}\right)\right),
\label{convar}
\eeqa
and so on.

The regimes of short samples ($N\ll N_\x$)
and of long samples ($N\gg N_\x$) deserve to be considered separately.

\subsection{Short samples}

We first consider relatively short samples,
such that $N\ll N_\x$, i.e., $X\ll1$.
In this regime, and for $\abs{E}<2$, the contour integral in~(\ref{conave})
is dominated by a saddle point at $s_c\approx(2-\abs{E})/x^{2H}\gg1$.
We thus obtain the following exponentially small estimate
\beq
\mean{\g_N}\sim\exp\left(-\frac{(2-\abs{E})^2}{2\,X^{2H}}\right)
\label{inst}
\eeq
for the mean effective Lyapunov exponent.

The right-hand side of~(\ref{inst}) has a simple interpretation.
It scales indeed as the probability that $E-V_N$ has reached the closest
forbidden zone, i.e., $E-V_N=2$ if $E>0$ and $E-V_N=-2$ if $E<0$,
so that the superlocalization phenomenon just sets in.

For a weak but finite disorder strength $\D$,
the mean effective Lyapunov exponent~$\mean{\g_N}$
is already of order $\D^2$ in the regime of usual localization,
i.e., for $X\ll1$.
The onset of superlocalization manifests itself as a crossover
of $\mean{\g_N}$ to the behavior~(\ref{inst}),
and it takes place for a system size such that
\beq
X\sim\abs{\ln\D}^{-1/(2H)}\ll1,
\qquad\hbox{i.e.,}\quad
N\sim\left(\D\abs{\ln\D}^{1/2}\right)^{-1/H}.
\eeq

\subsection{Long samples}

The superlocalization phenomenon is fully developed
in the converse regime of long samples ($N\gg N_\x$, i.e., $X\gg1$).
The contour integral in~(\ref{conave})
is then dominated by the branch cut of $K_0(2s)\approx-(\ln s+\C)$ at $s=0$,
with $\C$ being Euler's constant.
We thus obtain
\beq
\mean{\g_N}\approx H(\ln X-1)-\frac{\C+\ln 2}{2}.
\label{masy}
\eeq
The mean effective Lyapunov exponent
thus diverges logarithmically with the system size $N$ for $N\gg N_\x$,
irrespective of the energy $E$.

\begin{figure}[htb]
\begin{center}
\includegraphics[angle=90,width=.6\linewidth]{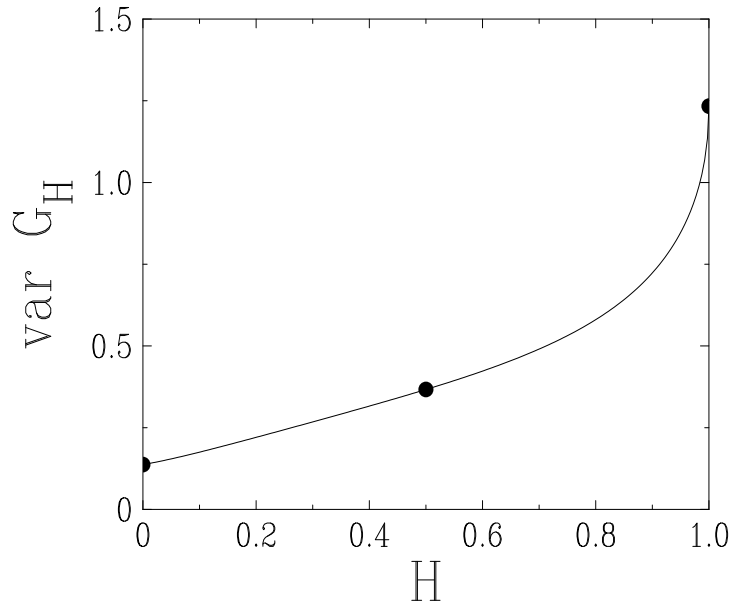}
\caption{\small
Plot of the variance of the functional $G_H$ entering the result~(\ref{gas}),
evaluated from the exact expression~(\ref{gvar}),
against the Hurst exponent $H$.
Symbols: results~(\ref{zer}), (\ref{gdemi}), and~(\ref{un})
for the particular cases $H=0$, $1/2$, and~$1$.}
\label{figvarg}
\end{center}
\end{figure}

\begin{figure}[htb]
\begin{center}
\includegraphics[angle=90,width=.6\linewidth]{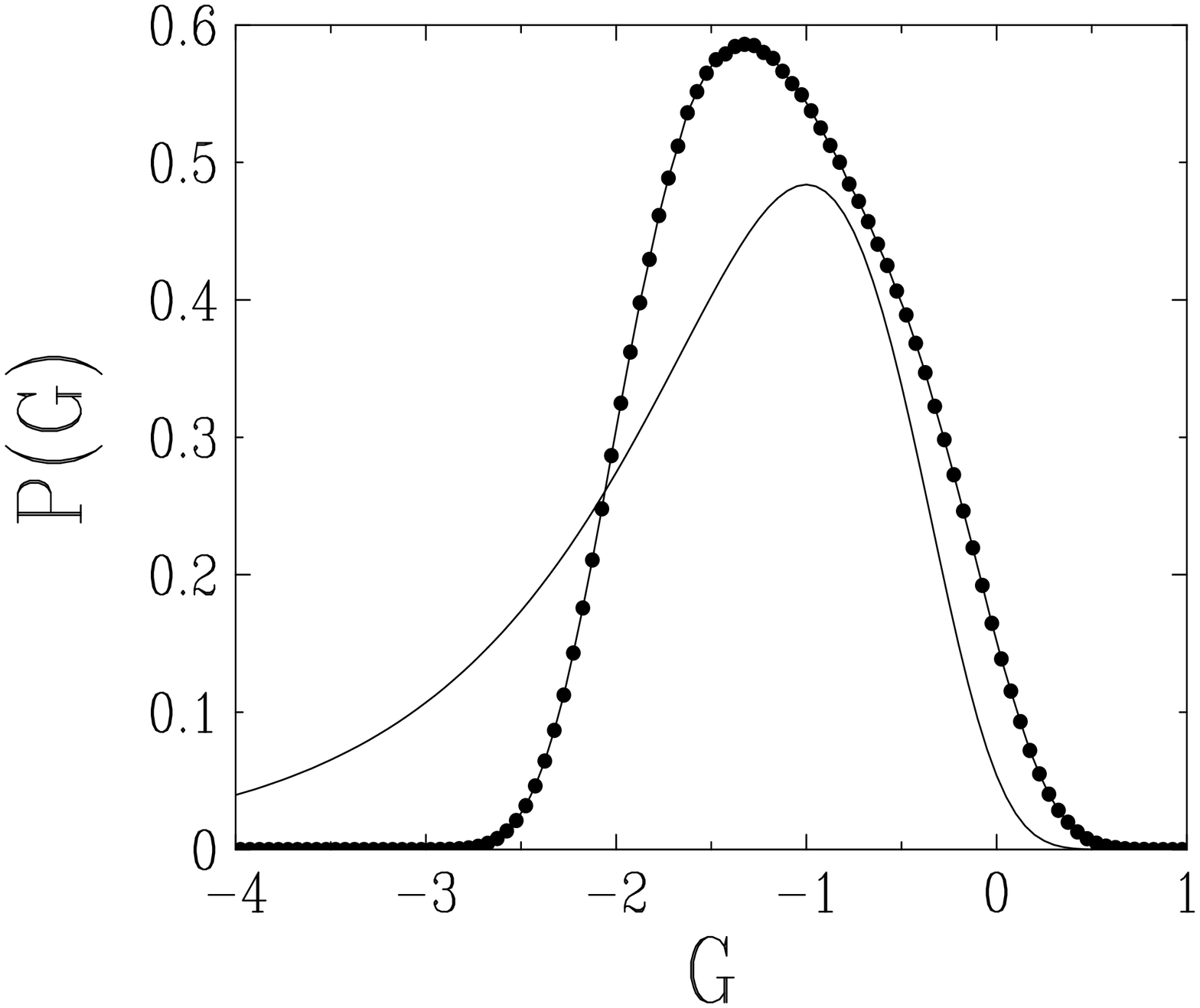}
\caption{\small
Plot of the probability density
of the functional $G_H$ entering the result~(\ref{gas}).
Thin full line: exact result~(\ref{ball}) for $H=1$ (ballistic limit).
Line with symbols (showing histogram bins):
numerical result for $H=1/2$ (Brownian case).}
\label{figg}
\end{center}
\end{figure}

As a matter of fact, the full distribution of the effective Lyapunov exponent
simplifies in the limit of a long sample.
Indeed, for $n\gg N_\x$ one has $\mean{V_n^2}\gg1$,
so that most often $\abs{V_n}\gg1$, hence $\Re t_n\approx\ln\abs{V_n}$, and
\beq
\g_N\approx\frac{1}{N}\sum_{n=0}^{N-1}\ln\abs{V_n}.
\eeq
The rescaling $x=n/N$ yields the following asymptotic form
of the effective Lyapunov exponent of long samples:
\beq
\g_N\approx H\ln X+G_H\approx H\ln(N/N_\x)+G_H\approx\ln(\D N^H)+G_H,
\label{gas}
\eeq
where the random additive term $G_H$ is the following non-linear functional
of the normalized fractional Brownian motion:
\beq
G_H=\int_0^1\d x\,\ln\abs{v(x)}.
\label{gfun}
\eeq
The expression~(\ref{gas}), which holds irrespectively of the energy $E$,
is the main result of this section.
The functional $G_H$ is investigated in the Appendix.
It has a universal distribution which only depends on the Hurst exponent $H$.
In the case $H=1/2$, corresponding to usual Brownian motion,
similar non-linear integral functionals have been met in several
physical problems~\cite{monmaj}.
To our knowledge, the distribution of $G_H$ is not known explicitly,
even in the Brownian case.
Expressions for the mean and the variance of $G_H$
are derived in the Appendix for all $H$,
as well as the full distribution of $G_1$ in the ballistic limit ($H=1$).
The mean $\mean{G_H}$ has a simple linear dependence~(\ref{gyave})
on the Hurst exponent~$H$, which can already be read off from~(\ref{masy}),
while the more complex expression~(\ref{gvar}) for the variance $\var{G_H}$
only simplifies for $H=0$, $1/2$, and~$1$.
Figure~\ref{figvarg} illustrates the dependence of $\var{G_H}$
on the exponent $H$,
whereas Figure~\ref{figg} shows a plot of the probability density
of $G_1$ (ballistic case: exact expression~(\ref{ball}))
and of $G_{1/2}$ (Brownian case),
the latter being measured from an ensemble of numerically generated,
and suitably rescaled, long random walks.
The distributions is observed to be clearly asymmetric (skew),
with very fast decaying tails at large values of $G_H$,
and at small values of $G_H$ for any $H<1$.

\subsection{Numerical results}

We have confronted our analytical predictions
with numerical results in the Brownian case ($H=1/2$).
This situation is indeed simpler to handle than the generic one.
On the one hand, the integrals over the rescaled spatial positions $x$, $y$
in the predictions~(\ref{conave}),~(\ref{convar}) are elementary.
On the other hand, sequences of Brownian potentials $V_n$ can easily
be generated numerically,
by summing independent increments $\eps_n=\pm\D$ with $\D\ll1$,
according to~(\ref{inc}), with $V_0=E=0$.

\begin{figure}[htb]
\begin{center}
\includegraphics[angle=90,width=.6\linewidth]{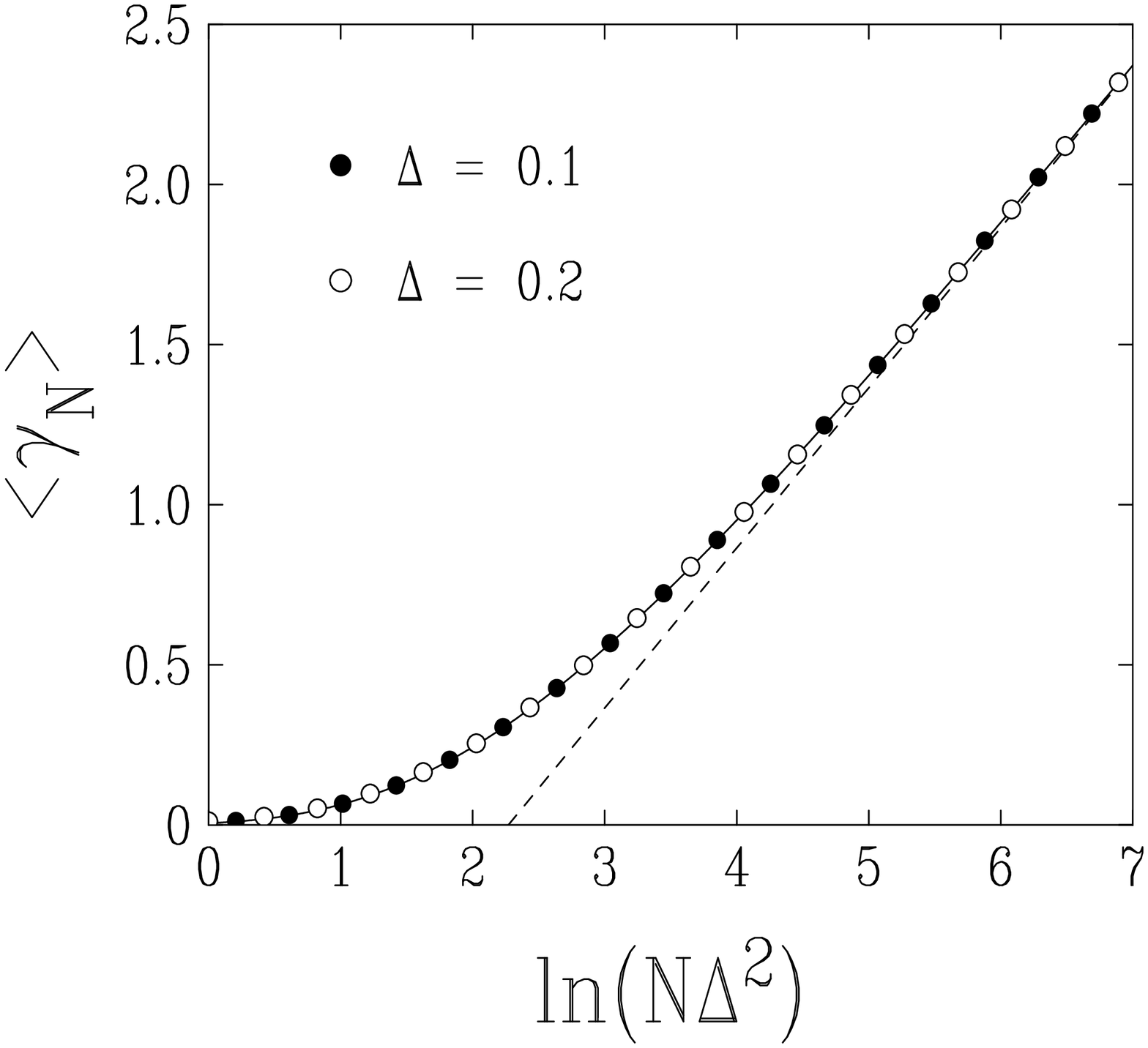}
\caption{\small
Plot of the mean effective Lyapunov exponent for a Brownian potential
($H=1/2$) with $E=0$,
against the logarithm of the scaling variable $X=N\D^2$.
Full curve: analytical prediction~(\ref{conave}).
Dashed line: asymptotic behavior~(\ref{masy}),
i.e., $\mean{\g_N}\approx(\ln(N\D^2)-1-\C-\ln 2)/2$.
Symbols: numerical data for $\D=0.1$ and $\D=0.2$.}
\label{fig1}
\end{center}
\end{figure}

\begin{figure}[htb]
\begin{center}
\includegraphics[angle=90,width=.6\linewidth]{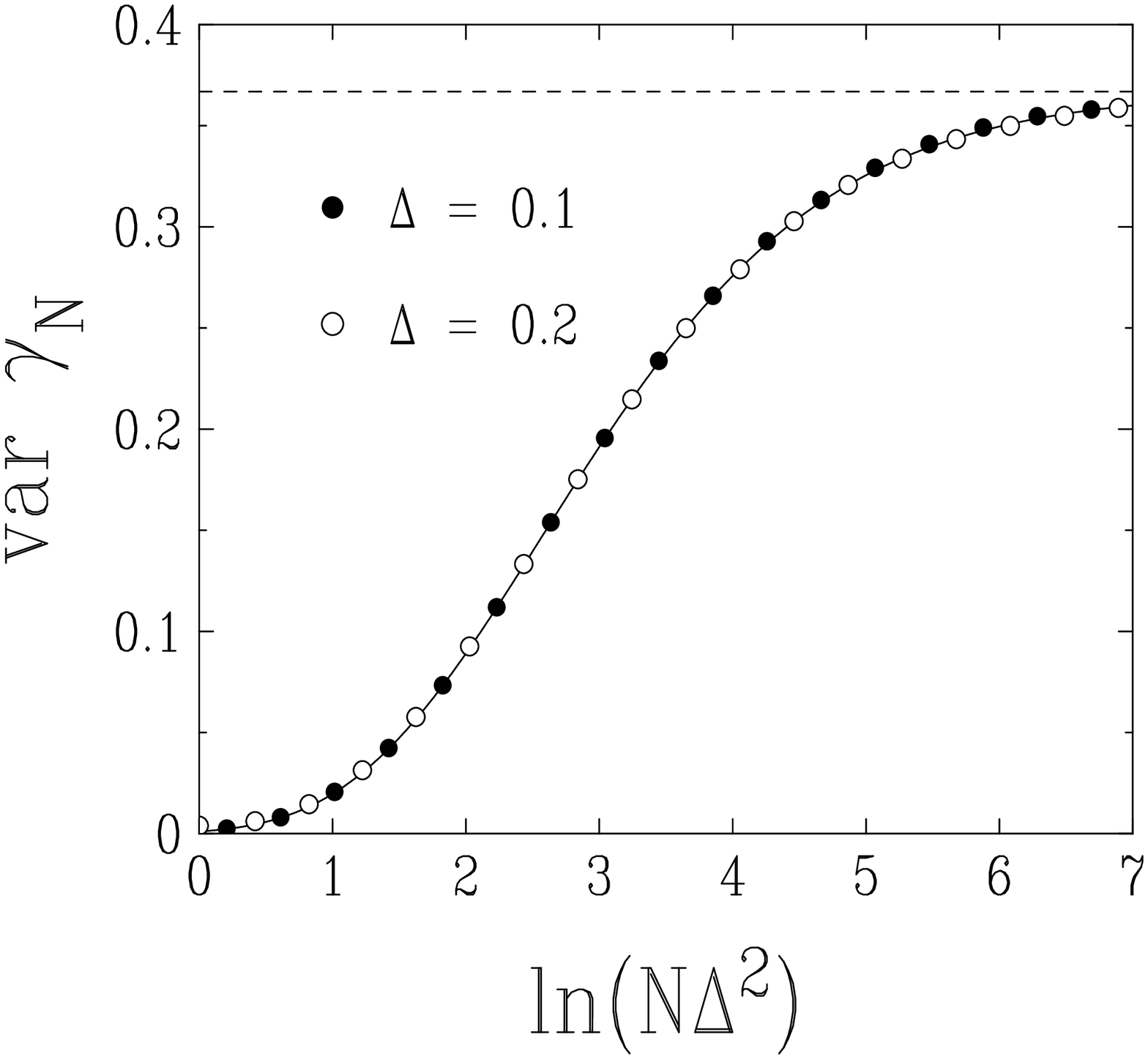}
\caption{\small
Plot of the variance of the effective Lyapunov exponent
for a Brownian potential ($H=1/2$) with $E=0$,
against the logarithm of the scaling variable $X=N\D^2$.
Full curve: analytical prediction~(\ref{convar}).
Dashed line: limiting value~(\ref{gdemi}):
$\var{G_{1/2}}=(\pi^2-4)/16=0.366850$.
Symbols: numerical data for $\D=0.1$ and $\D=0.2$.}
\label{fig2}
\end{center}
\end{figure}

The growing solution $\psi_n$ has been evaluated
by means of the recursion~(\ref{ricrec}) for the Riccati variables $R_n$,
with a random initial condition ($R_1=\tan\phi$ with a uniform angle~$\phi$).
The effective Lyapunov exponent has been measured by means of~(\ref{garic}).
The mean and the variance of the effective Lyapunov exponent
are respectively shown in Figures~\ref{fig1} and~\ref{fig2}.
The continuous curves represent
the analytical results~(\ref{conave}),~(\ref{convar}),
whereas symbols show numerical data for $10^5$ samples
with two strengths of disorder, $\D=0.1$ and $\D=0.2$.
The accuracy of the data collapse and of the quantitative agreement
with theoretical predictions provides a convincing check of our analysis.

\section{The continuum Schr\"odinger equation}

We now turn to the Schr\"odinger equation
in a one-dimensional potential $V(x)$, which reads
\beq
-\psi''(x)+V(x)\psi(x)=E\psi(x),
\label{sch}
\eeq
in reduced units.
The potential $V(x)$ is again assumed to be a weak self-affine Gaussian
disordered potential with Hurst exponent $0<H<1$, such that
\beq
\mean{(V(x)-V(y))^2}=\D^2\abs{x-y}^{2H},
\label{chur}
\eeq
with $V(0)=0$ and $\D^2\ll1$.

\subsection{General results}

The present situation shares most of the characteristic features
of the tight-binding problem investigated in Section~2.
Its analysis will therefore only be described succinctly.

A long enough sample is again generically an alternation
of classically allowed zones ($V(x)<E$) and of forbidden zones ($V(x)>E$).
We assume $E>0$, so that the origin belongs to an allowed zone.
The growth rate of the generic solution of~(\ref{sch})
up to distance~$L$ is measured by the effective Lyapunov exponent
\beq
\g(L)=\frac{1}{L}\ln\abs{\psi(L)},
\label{lys}
\eeq

The local growth rate of wavefunctions
is again proportional to $\D^2$ in allowed zones,
whereas it is of order unity in forbidden zones.
It is therefore legitimate to use the celebrated W.K.B.
integral (see e.g.~\cite{ll,wkb})
\beq
\psi(x)\sim\exp\left(\int_0^x\d y\,\sqrt{(V(y)-E)_+}\right),
\label{psiwkb}
\eeq
with the definition
\beq
\sqrt{(V(y)-E)_+}=\left\{\matrix{
\sqrt{V(y)-E}\hfill&\hbox{in forbidden zones}&(V(y)>E),\hfill\cr\cr
0\hfill&\hbox{in allowed zones}\hfill&(V(y)<E).\hfill
}\right.
\label{dichoc}
\eeq

The integral formula~(\ref{psiwkb}) provides a quantitative description,
with exponential accuracy,
of the growing solution to~(\ref{sch}) in the forbidden zones of the potential,
whenever the latter is slowly varying.
One has therefore
\beq
\g(L)\approx\frac{1}{L}\int_0^L\d x\,\sqrt{(V(x)-E)_+}.
\label{lyc}
\eeq

Equations~(\ref{lyc}) and~(\ref{dichoc}) are the continuum analogues
of~(\ref{lyd}) and~(\ref{dichod}).
They constitute the starting point of the subsequent
analysis of the statistics of the effective Lyapunov exponent.
It is again advantageous to introduce a Laplace representation:
\beq
\sqrt{(V(x)-E)_+}=\int\!\frac{\d s}{2\i\pi}\,\sqrt{\frac{\pi}{4s^3}}
\,\exp(s(V(x)-E)).
\eeq
The correlation functions of the variables $\sqrt{(V(x)-E)_+}$,
and therefore the moments of the effective Lyapunov exponent,
can again be expressed as explicit multiple integrals.
We will only need the expression of the mean effective Lyapunov exponent,
which reads
\beq
\mean{\g(L)}=\frac{1}{L}\int_0^L\d x
\int\!\frac{\d s}{2\i\pi}\,\sqrt{\frac{\pi}{4s^3}}
\exp\left(\demi\D^2s^2x^{2H}-sE\right).
\eeq
The rescalings $z=x/\ell_\x$, $u=sE$ show that the above result scales as
\beq
\mean{\g(L)}=\frac{E^{1/2}}{X}\int_0^X\d z
\int\!\frac{\d u}{2\i\pi}\,\sqrt{\frac{\pi}{4u^3}}
\exp\left(\demi u^2z^{2H}-u\right),
\label{schave}
\eeq
in terms of the length ratio
\beq
X=\frac{L}{\ell_\x},
\eeq
where the energy-dependent crossover length
\beq
\ell_\x=(E/\D)^{1/H}
\eeq
is the typical size of the first allowed zone containing the origin.

\subsection{Short samples}

For rather small samples, such that $L\ll\ell_\x$, i.e., $X\ll1$,
the contour integral in~(\ref{schave})
is dominated by a saddle point at $u_c\approx z^{-2H}\gg1$.
We are thus left with the exponentially small estimate
\beq
\mean{\g(L)}\sim\exp\left(-\frac{1}{2\,X^{2H}}\right).
\eeq
This expression again scales as the probability that $V(L)$ is equal to $E$,
so that a forbidden zone has just been entered.

For a weak but finite disorder strength $\D$,
the mean effective Lyapunov exponent
again has a finite limit of order $\D^2$ for $X\ll1$.
The onset of superlocalization again takes place for a system size such that
\beq
X\sim\abs{\ln\D}^{-1/(2H)}\ll1,
\qquad\hbox{i.e.,}\quad
L\sim\left(\frac{E}{\D\abs{\ln\D}^{1/2}}\right)^{1/H}.
\eeq

\subsection{Long samples}

The superlocalization phenomenon is fully developed
in the converse regime of long samples ($L\gg\ell_\x$, i.e., $X\gg1$).
The contour integral in~(\ref{schave})
is dominated by the square-root branch cut at $s=0$.
We thus obtain a power-law growth of the mean effective Lyapunov exponent:
\beq
\mean{\g(L)}\approx\frac{2^{1/4}}{(H+2)\sqrt{\pi}}
\,\Gamma\!\left(\tquart\right)(\D L^H)^{1/2},
\label{meany}
\eeq
irrespective of the energy $E$.

\begin{figure}[htb]
\begin{center}
\includegraphics[angle=90,width=.6\linewidth]{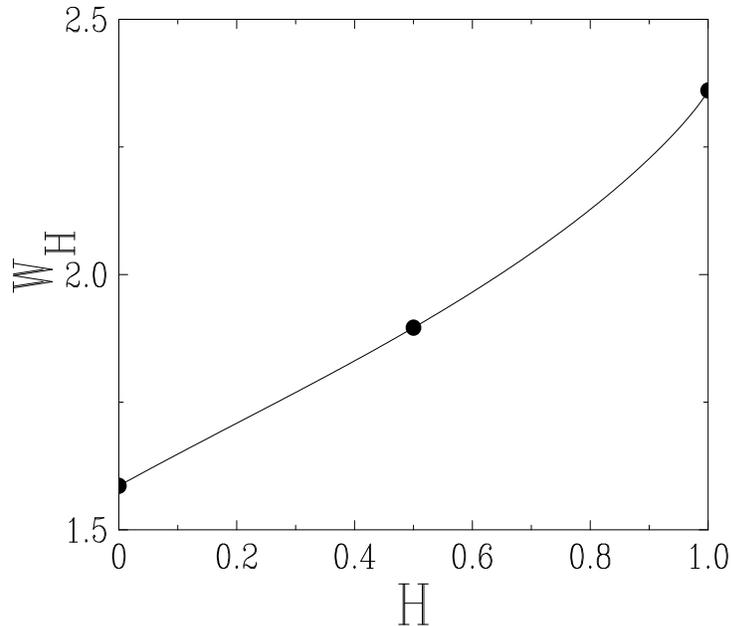}
\caption{\small
Plot of the moment ratio $W_H$, defined in~(\ref{whdef}),
of the quantity $Y_H$ entering the result~(\ref{glas}),
evaluated from the exact expression~(\ref{yw}),
against the Hurst exponent $H$.
Symbols: results~(\ref{zer}), (\ref{wdemi}), and~(\ref{un})
for the particular cases $H=0$, $1/2$, and $1$.}
\label{figwh}
\end{center}
\end{figure}

\begin{figure}[htb]
\begin{center}
\includegraphics[angle=90,width=.6\linewidth]{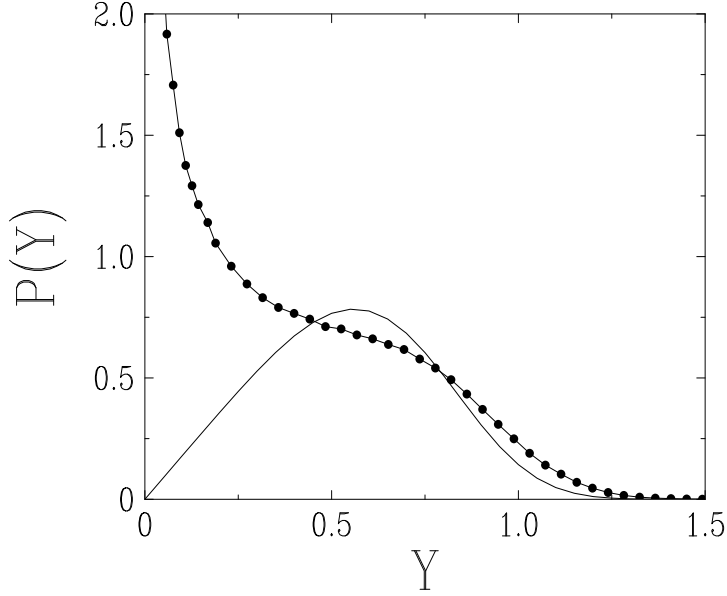}
\caption{\small
Plot of the probability density
of the random variable $Y_H$ entering the result~(\ref{glas}).
Thin full line: continuous component (delta peak at $Y=0$ omitted)
of the exact result~(\ref{ball}) for $H=1$ (ballistic limit).
Line with symbols (showing histogram bins):
numerical result for $H=1/2$ (Brownian case).}
\label{figy}
\end{center}
\end{figure}

The full distribution of the effective Lyapunov exponent
again simplifies in the limit of a long sample.
Indeed, for $x\gg\ell_\x$, $E$ is most often negligible with respect to $V(x)$.
Equation~(\ref{lyc}) therefore simplifies to
\beq
\g(L)\approx\frac{1}{L}\int_0^L\d x\,\sqrt{(V(x))_+}.
\eeq
The rescaling of $x$ by $L$ yields the following asymptotic form
of the effective Lyapunov exponent of long samples:
\beq
\g(L)\approx E^{1/2}\,X^{H/2}\,Y_H\approx E^{1/2}\,(L/\ell_\x)^{H/2}\,Y_H
\approx(\D L^H)^{1/2}\,Y_H,
\label{glas}
\eeq
where the fluctuating factor $Y_H$ is another non-linear functional
of the normalized fractional Brownian motion:
\beq
Y_H=\int_0^1\d x\,\sqrt{(v(x))_+}.
\label{yfun}
\eeq
This functional is investigated in the Appendix.
It has a universal distribution which only depends on the Hurst exponent $H$.
Its first two moments are evaluated for all values of $H$
[see respectively~(\ref{gyave}) and~(\ref{y2})],
as well as the full distribution of $Y_1$ in the ballistic limit.
Figure~\ref{figwh} illustrates the dependence of the dimensionless ratio
\beq
W_H=\frac{\mean{Y_H^2}}{\mean{Y_H}^2}
\label{whdef}
\eeq
on the exponent $H$.
Figure~\ref{figy} shows a plot of the probability density
of $Y_1$ (ballistic case: exact expression~(\ref{ball}))
and of $Y_{1/2}$ (Brownian case),
the latter being measured from an ensemble of numerically generated,
and suitably rescaled, long random walks.
The delta peak at $Y=0$, which is present in the ballistic limit ($H=1$),
gets smeared out into a continuous density which diverges as $Y\to0$
for any $H<1$.

\section{Discussion}

The present work is devoted to the specific features of localization
in a weak self-affine disordered potential in one dimension,
whose fluctuations grow with distance with a positive Hurst exponent $H$.
The most salient feature of such non-stationary potentials
is that the effective disorder strength depends on the spatial scale.
Even in the regime of most physical interest,
where the fluctuations of the disordered potential are small
at the microscopic scale ($\D\ll1$), the effective disorder becomes strong
beyond a well-defined crossover length $N_\x$ or $\ell_\x$,
which diverges as $\D^{-1/H}$.
Samples much longer than this crossover length
typically consist of an alternation of classically allowed zones,
where the growth rate of a generic wavefunction is small
and proportional to the square disorder strength~$\D^2$,
and of forbidden zones, where this growth rate is large.
Typical eigenstates on such samples turn out to be superlocalized~\cite{b2,b3}.

In this paper we present a parallel analytical investigation
of the statistics of superlocalized eigenstates,
for both the discrete tight-binding model
and the continuum Schr\"odinger equation.
First of all, the clear separation of the growth rates of wavefunctions
in allowed and in forbidden zones in the weak-disorder regime
is shown to allow for a semi-classical treatment
of the superlocalization phenomenon, by means of the W.K.B. formalism.
The key quantity is the size-dependent effective Lyapunov exponent,
which is closely related to the transmission across the sample,
and therefore to its zero-temperature conductance by the Landauer formula.

For the discrete tight-binding model, investigated in Section~2,
the effective Lyapunov exponent behaves as
\beq
\g_N\approx H\ln(N/N_\x)+G_H\approx\ln(\D N^H)+G_H
\label{dgas}
\eeq
in the fully superlocalized regime of long samples
($N\gg N_\x$) [see~(\ref{gas})].
The fluctuating part~$G_H$ is an {\it additive term} of order unity.
It is given by the functional~(\ref{gfun}),
whose limit distribution only depends on the Hurst exponent $H$.
These results translate as follows in terms of physical quantities.
One of the most characteristic features of superlocalization
is the fall-off of the effective localization length $\xi_N$
as a function of $N$.
This decay is however only logarithmic, as $\xi_N\approx1/[H\ln(N/N_\x)]$.
As far as the transmission $T_N$ is concerned
(or equivalently the conductance $g_N$),
the relevant quantity is its logarithm, as in most localization problems.
Its mean grows as $\mean{\ln T_N}\approx-2HN\ln(N/N_\x)$,
whereas its variance grows as $\var{\ln T_N}\approx4\var G_H N^2$.
The logarithm of the transmission is therefore
{\it marginally self-averaging},
as its relative variance $\var{\ln T_N}/\mean{\ln T_N}^2\sim1/[\ln(N/N_\x)]^2$
falls off logarithmically.

For the continuum Schr\"odinger equation, investigated in Section~3,
the effective Lyapunov exponent behaves as
\beq
\g(L)\approx E^{1/2}\,(L/\ell_\x)^{H/2}\,Y_H\approx(\D L^H)^{1/2}\,Y_H
\label{dglas}
\eeq
[see~(\ref{glas})] in the fully superlocalized regime.
The fluctuating part $Y_H$ of this result, given by the functional~(\ref{yfun}),
is now involved {\it as a multiplicative factor.}
This multiplicative law generates more pronounced superlocalization effects,
and especially stronger fluctuations in physical quantities.
The effective localization length $\xi(L)\approx1/(Y_H\D^{1/2}L^{H/2})$
now decays as a power of the sample length, with a fluctuating amplitude.
As far as the transmission is concerned,
its mean grows as $\mean{\ln T(L)}\sim-\D^{1/2}L^{1+H/2}$,
i.e., more rapidly than linearly with the sample length.
The logarithm of the transmission is now {\it fully non-self-averaging},
as its relative variance $\var{\ln T(L)}/\mean{\ln T(L)}^2\to W_H-1$
has a non-trivial limit for $L\gg\ell_\x$.

The effective Lyapunov exponent and related physical quantities
therefore exhibit two different kinds of statistics
in the discrete and in the continuum formalism.
The basic difference between both situations
is already present in the case of a constant potential.
It is indeed deeply rooted in the dispersion relations of the models.
For the tight-binding model with a constant site potential~$V$ outside the band,
the growth rate~$t$ of wavefunctions
is such that $E-V=2\cosh t$, and therefore only diverges logarithmically
as $t\approx\ln\abs{V}$ for large $V$.
For the continuum Schr\"odinger equation with a constant potential~$V$,
the growth rate now obeys $V-E=t^2$, and diverges as $t\approx\sqrt{V}$.
Replacing in the above dispersion estimates $t$ by $\g$,
and $V$ by the product of a random number of order unity by
$V_\rms\sim\D N^H$ or $\D L^H$, leads to the correct qualitative forms
of the scaling equations~(\ref{dgas}) and~(\ref{dglas}).
Finally, this line of reasoning also shows
that the results for the Schr\"odinger equation cannot be recovered
from those of the tight-binding model by going to the continuum limit,
because this limit corresponds to $t\to0$,
whereas superlocalization is essentially due to large values of $t$.
Table~\ref{one} summarizes the discussion.

\begin{table}[ht]
\begin{center}
\begin{tabular}{|c|c|c|c|c|}
\hline
Quantity&tight-binding model&Schr\"odinger equation&Anderson\\
&$N\gg N_\x$ sites&length $L\gg\ell_\x$&localization\\
\hline
$\mean{\xi}$&$1/\ln(N/N_\x)$&$(\D L^H)^{-1/2}$&$1/\D^2$\\
$-\mean{\ln T}$&$N\ln(N/N_\x)$&$\D^{1/2}L^{1+H/2}$&$\D^2 L$\\
$\var{\ln T}/\mean{\ln T}^2$&$1/[\ln(N/N_\x)]^2$&constant&$1/(\D^2L)$\\
\hline
\end{tabular}
\caption{\small Scaling behavior of various quantities
in the regime of fully developed superlocalization:
mean effective localization length, mean and relative variance
of the logarithm of the transmission (conductance).
Only the qualitative scaling behavior is given
(dependence on the sample size $N$ or $L$ and on the disorder strength $\D$).
The emphasis is put on the difference between
the discrete tight-binding model
(additive fluctuation in the effective Lyapunov exponent)
and the continuum Schr\"odinger equation (multiplicative fluctuation).
The well-known results for conventional Anderson localization
are recalled for comparison.}
\label{one}
\end{center}
\end{table}

Let us close up with a word on the generality of our results.
For any weak self-affine (not necessarily Gaussian)
random potential obeying power laws
of the form~(\ref{hur}) or~(\ref{chur}), with any positive Hurst exponent~$H$,
both the functional form of the results~(\ref{dgas}),~(\ref{dglas})
and the scaling laws recalled in Table~\ref{one}
can be shown to still hold true.
The fluctuating parts $G_H$ and~$Y_H$ are, however,
only given by~(\ref{gfun}) and~(\ref{yfun}) for the Gaussian
self-affine potentials associated with fractional Brownian motion.
Their distribution involve in general further details
of the statistics of the potentials.

\subsubsection*{Acknowledgements}

It is a pleasure to thank Dominique Boos\'e for very stimulating discussions.

\appendix
\setcounter{equation}{0}
\def\theequation{A.\arabic{equation}}
\section*{Appendix. The functionals $G_H$ and $Y_H$}

This Appendix is devoted to the random quantities $G_H$ and $Y_H$,
which respectively enter our key results~(\ref{gas}) and~(\ref{glas}).
They are defined as the following non-linear functionals
\beq
G_H=\int_0^1\d x\,\ln\abs{v(x)},\qquad
Y_H=\int_0^1\d x\,\sqrt{(v(x))_+},
\eeq
of the normalized fractional Brownian motion $v(x)$
with Hurst exponent $0<H<1$ on $0\le x\le1$, so that $v(0)=0$ and
\beq
\mean{(v(x)-v(y))^2}=\abs{x-y}^{2H}.
\eeq

\subsubsection*{First moments}

The means (first moments)
\beq
\mean{G_H}=\int_0^1\d x\,\mean{\ln\abs{v(x)}},\qquad
\mean{Y_H}=\int_0^1\d x\,\bigmean{\sqrt{(v(x))_+}},
\eeq
can be readily evaluated.
Indeed, $v(x)$ is a Gaussian with variance
$\mean{v(x)^2}=x^{2H}$, so that its probability density reads
\beq
P(v(x))=\frac{1}{x^H\sqrt{2\pi}}\,\exp\left(-\frac{v(x)^2}{2\,x^{2H}}\right).
\eeq
Elementary integrals yield
\beq
\mean{\ln\abs{v(x)}}=H\ln x-\frac{\C+\ln 2}{2},\qquad
\bigmean{\sqrt{(v(x))_+}}
=\frac{1}{2^{3/4}\sqrt{\pi}}\,\Gamma\!\left(\tquart\right)x^{H/2},
\eeq
so that
\beq
\mean{G_H}=-H-\frac{\C+\ln 2}{2},
\qquad
\mean{Y_H}=\frac{2^{1/4}}{(H+2)\sqrt{\pi}}\,\Gamma\!\left(\tquart\right).
\label{gyave}
\eeq

\subsubsection*{Second moments}

The second moments
\beq
\mean{G_H^2}=2\int_0^1\d x\int_x^1\d y\,\mean{\ln\abs{v(x)}\,\ln\abs{v(y)}},
\quad
\mean{Y_H^2}=2\int_0^1\d x\int_x^1\d y\,\bigmean{\sqrt{(v(x))_+(v(y))_+}},
\eeq
now involve two-point observables,
expressed in terms of the correlated Gaussian variables $v(x)$ and $v(y)$,
such that
\beq
\mean{v(x)^2}=x^{2H},\qquad\mean{v(y)^2}=y^{2H},\qquad
\mean{v(x)v(y)}=\demi\left(x^{2H}+y^{2H}-(y-x)^{2H}\right).
\label{vxy}
\eeq
It is convenient to use polar co-ordinates in the $v(x)$, $v(y)$ plane.
For $x\le y$, we set
\beq
z=\frac{x}{y}\le1,
\eeq
and
\beq
v(x)=r\cos\theta,\qquad v(y)=\frac{r}{z^H}\,\sin(\theta+\alpha),
\eeq
with
\beq
\sin\alpha
=\frac{\mean{v(x)v(y)}}{\sqrt{\mean{v(x)^2}\,\mean{v(y)^2}}}
=\frac{1+z^{2H}-(1-z)^{2H}}{2z^H}\qquad(0\le\alpha\le\pi/2).
\label{aldef}
\eeq
This parametrization ensures that the angle $\theta$ is
uniformly distributed between $0$ and $2\pi$,
whereas the radial variable~$r$ has a probability density
\beq
P_r(r)=\frac{r}{x^{2H}}\,\exp\!\left(-\frac{r^2}{2\,x^{2H}}\right).
\eeq

The calculation of $\mean{\ln\abs{v(x)}\,\ln\abs{v(y)}}$
splits into the radial integrals
\beq
\mean{\ln r}=\frac{\ln(2x^{2H})-\C}{2},\qquad
\mean{(\ln r)^2}_\c=\frac{\pi^2}{24},
\label{rad}
\eeq
where $\mean{\cdots}_\c$ denotes the connected part, and the angular integrals
\beq
\mean{\ln\abs{\!\cos\theta}}=\mean{\ln\abs{\!\sin\theta}}=-\ln2,\qquad
\mean{\ln\abs{\!\cos(\theta)}\,\ln\abs{\!\sin(\theta+\alpha)}}_\c=
\frac{\alpha^2}{2}-\frac{\pi^2}{24}.
\label{ang}
\eeq
The derivation of the second result requires the use of the Fourier series
\beq
\sum_{n=1}^\infty(-1)^n\,\frac{\cos(2n\theta)}{n}=-\ln(2\abs{\!\cos\theta}),
\qquad
\sum_{n=1}^\infty(-1)^n\,\frac{\cos(2n\theta)}{n^2}=\theta^2-\frac{\pi^2}{12},
\eeq
where the latter equality holds for $\abs{\theta}\le\pi$.
We thus obtain the following simple expression
\beq
\mean{\ln\abs{v(x)}\,\ln\abs{v(y)}}_\c=\frac{\alpha^2}{2}
\eeq
for the connected correlation function,
which leads us to the result
\beq
\var{G_H}=\mean{G_H^2}_\c=\demi\int_0^1\alpha^2\,\d z,
\label{gvar}
\eeq
where $\alpha$ is defined in terms of $z$ by~(\ref{aldef}).

Similarly, the calculation of $\bigmean{\sqrt{(v(x))_+(v(y))_+}}$
splits into the radial integral
\beq
\mean{r}=\sqrt{\frac{\pi}{2}}\,x^H,
\eeq
and the angular integral
\beq
\int_{-\alpha}^{\pi/2}
\frac{\d\theta}{2\pi}\sqrt{\cos\theta\,\sin(\theta+\alpha)}
=\frac{k^2}{\pi}\int_0^{\pi/2}\frac{\cos^2\phi\,\d\phi}{\sqrt{1-k^2\sin^2\phi}}
=\frac{1}{\pi}\left(\E(k)-(1-k^2)\K(k)\right),
\label{angul}
\eeq
where we have set
\beq
\sin\left(\theta+\frac{\alpha}{2}-\frac{\pi}{4}\right)=k\sin\phi,
\eeq
with
\beq
k^2=\sin^2\left(\frac{\pi}{4}+\frac{\alpha}{2}\right)
=\frac{1+\sin\alpha}{2}=\frac{(1+z^H)^2-(1-z)^{2H}}{4z^H},
\label{kdef}
\eeq
and where $\E(k)$ and $\K(k)$ are complete elliptic integrals.
Finally,
\beq
\mean{Y_H^2}=\frac{1}{H+2}\sqrt{\frac{2}{\pi}}
\int_0^1\left(\E(k)-(1-k^2)\K(k)\right)z^{H/2}\,\d z,
\label{y2}
\eeq
so that
\beq
W_H=\frac{\mean{Y_H^2}}{\mean{Y_H}^2}
=\frac{H+2}{2\,\pi^{3/2}}\,\Gamma\!\left(\quart\right)^2
\int_0^1\left(\E(k)-(1-k^2)\K(k)\right)z^{H/2}\,\d z,
\label{yw}
\eeq
where $k$ is defined in terms of $z$ by~(\ref{kdef}).

The general expressions~(\ref{gvar}) and~(\ref{yw})
simplify in the following particular cases.

\subsubsection*{Ultraslow limit ($H=0$)}

\noindent In this limit we have
\beq
\alpha=\frac{\pi}{6},\qquad k=\frac{\sqrt{3}}{2},
\eeq
irrespective of $z$, hence
\beq
\var{G_0}=\frac{\pi^2}{72}=0.137078,\qquad
W_0=\frac{1}{\pi^{3/2}}\,\Gamma\!\left(\quart\right)^2
\left(\E\left(\frac{\sqrt{3}}{2}\right)
-\quart\,\K\left(\frac{\sqrt{3}}{2}\right)\right)=1.586206.
\label{zer}
\eeq

\subsubsection*{Brownian case ($H=1/2$)}

\noindent In this case we have
\beq
\sin\alpha=2k^2-1=\sqrt{z},
\eeq
so that a direct integration yields
\beq
\var{G_{1/2}}=\frac{\pi^2-4}{16}=0.366850.
\label{gdemi}
\eeq
In order to evaluate $W_{1/2}$,
it is more convenient to go back to the middle expression of~(\ref{angul}),
and to integrate first over $k$, then over $\phi$.
We thus obtain after some algebra
\beq
W_{1/2}=\frac{5}{288\pi^{3/2}}\,\Gamma\!\left(\quart\right)^2
\left(50-3\sqrt{2}\,\ln\left(1+\sqrt{2}\right)\right)=1.895949.
\label{wdemi}
\eeq

\subsubsection*{Ballistic limit ($H=1$)}

\noindent In this case we have
\beq
\alpha=\frac{\pi}{2},\qquad k=1,
\eeq
irrespective of $z$, hence
\beq
\var{G_1}=\frac{\pi^2}{8}=1.233701,\qquad
W_1=\frac{1}{\pi^{3/2}}\,\Gamma\!\left(\quart\right)^2=2.360681.
\label{un}
\eeq
The constancy of $\alpha=\pi/2$
means that $v(x)$ and $v(y)$ are fully correlated.
Their correlation functions~(\ref{vxy})
indeed saturate the Cauchy-Schwarz inequality.

The whole fractional Brownian process degenerates into
a one-variable problem in the ballistic limit.
We have indeed
\beq
v(x)=xw,
\eeq
where $w\equiv v(1)$ is a Gaussian variable such that $\mean{w^2}=1$, hence
\beq
G_1=\ln\abs{w}-1,\qquad Y_1=\frac{2}{3}\sqrt{w_+}.
\eeq
By performing the appropriate changes of variables on the Gaussian law of $w$,
we get the following explicit expressions for the probability densities:
\beq
P(G_1)=\sqrt{\frac{2}{\pi}}\,\exp\left(G_1+1-\demi\,\e^{2(G_1+1)}\right),
\qquad
P(Y_1)=\demi\delta(Y_1)
+\frac{9\,Y_1}{2\sqrt{2\pi}}\,\exp\left(-\frac{81\,Y_1^4}{32}\right).
\label{ball}
\eeq


\end{document}